\documentclass[prc,twocolumn,showpacs,floatfix,aps,10pt]{revtex4-1}

\usepackage{amsmath}
\usepackage{graphicx}

\begin{document}

\title{
Decay and structure of the Hoyle state
}
\author{S.\ Ishikawa}  \email[E-mail:]{ishikawa@hosei.ac.jp}
\affiliation{ 
Science Research Center, Hosei University, 
2-17-1 Fujimi, Chiyoda, Tokyo 102-8160, Japan
} 

\date{\today}

\begin{abstract}
The first $0^+$ resonant state of the $^{12}$C nucleus ${}^{12}$C$(0_2^+)$, so called the Hoyle state, is investigated in a three-$\alpha$-particle (3-$\alpha$) model.
A wave function for the photodisintegration reaction of a $^{12}$C bound state to 3-$\alpha$ final states is defined and calculated by the Faddeev three-body formalism, in which three-body bound- and continuum states are treated consistently. 
From the wave function at the Hoyle state energy, I calculated distributions of outgoing $\alpha$-particles and density distributions at interior region of the Hoyle state. 
Results show that a process through a two-$\alpha$ resonant state is dominant in the decay and contributions of the rest process are very small, less than 1 \%. 
There appear  some peaks in the interior density distribution corresponding to configurations of an equilateral- and  an isosceles triangles.  
It turns out that these results are obtained independently of the choice of $\alpha$-particle interaction models, when they are made to reproduce the Hoyle state energy.
\end{abstract}

%

\pacs{
21.45.-v,  
25.70.Ef,  
27.20.+n  
}

\maketitle

{\em Introduction.} 
The Hoyle state \cite{Ho54} is a resonant state of the $^{12}$C nucleus at an energy just above the 3-$\alpha$ threshold, which  decays mainly to 3-$\alpha$ continuum states with a very small branching ratio of radiative decays to  $^{12}$C bound states  \cite{Aj90}.
Because of the existing of two-$\alpha$-particle resonant state ${}^8$Be$(0_1^+)$ 
($E[{}^{8}$Be$(0_1^+)] = 0.092$ MeV and a decay width $\Gamma_{\alpha\alpha}=5.57(25)$ eV \cite{Ti04}), the 3-$\alpha$ decay is dominated by a successive  process  being referred to as the sequential decay (SD) \cite{Fr94,Ma12,Ki12,Ra13,It14}, 
\begin{eqnarray}
{}^{12}\mathrm{C}(0_2^+) &\to& {}^{8}\mathrm{Be}(0_1^+) +  \alpha 
\cr
&\to&  \alpha +  \alpha +  \alpha. 
\end{eqnarray} 
This is a key feature in  evaluating the thermal nuclear reaction rate of the triple-alpha (3$\alpha$) process, by which three $\alpha$-particles are fused into a $^{12}$C nucleus in stars \cite{An99}.

On the other hand, the structure of the Hoyle state has been one of long-standing issues to study in Nuclear Physics. 
Some calculations show that the Hoyle state has a component consisting of  three $\alpha$-particles taking a certain geometric configuration, such as a linear chain, an equilateral triangle, or an isosceles triangle \cite{Mo56,Ue77,Ch07,Ep12,Ka07}. 
Since these calculations were performed essentially by an approximation that particles are confined in a limited volume, it is not clear how they decay from the resonant state at long distances. 
This leads to a requirement of proper treatments of three-body continuum states. 
In Ref. \cite{Is13}, I calculated  the 3$\alpha$ reaction rate by considering the inverse reaction of the  fusion, namely the  E2-photodisintegration of ${}^{12}$C$(2_1^+)$ state,
\begin{equation}
{}^{12}{\mathrm{C}}(2_1^+) + \gamma \to \alpha + \alpha + \alpha,
\label{eq:photo-dis-C}
\end{equation}
where the total angular momentum of the final 3-$\alpha$ state is 0.
There, a wave function for the reaction (\ref{eq:photo-dis-C}) is defined and solved by applying the Faddeev three-body formalism in coordinate space \cite{Fa61}. 
The calculated cross section as a function of the photon energy has a sharp peak corresponding to the Hoyle state as shown in Fig. 2 of Ref. \cite{Is13}.  
The solution provides not only a breakup amplitude to give the cross section, but also a 3-$\alpha$ wave function, from which the interior structure of 3-$\alpha$ system can be considered.  
Although the reaction (\ref{eq:photo-dis-C}) to populate the Hoyle state is not same as ones in previous experimental works to study the decay of the Hoyle state:  
inelastic reactions of the ${}^{12}\mathrm{C}$ ground state \cite{Fr94,Ma12,Ra13,It14,Ra11} or a transfer reactions \cite{Ki12}, 
once the long-lived state is formed, (see the narrow 3-$\alpha$ decay width $\Gamma_{3\alpha}$ in Table \ref{tab:3alphaP-parameters} below), 
the decay process  is expected to occur in common irrespective of the formation process. 
In this paper, therefore I will analyze the reaction (\ref{eq:photo-dis-C}) at the Hoyle state energy, and study 3-$\alpha$ decay modes as well as the density distribution of the three $\alpha$-particles at smaller distances.
In the following, after describing theoretical methods and models used in this work, I will report results of the calculations.

{\em Theoretical method.}
Let us consider a disintegration of ${}^{12}$C bound state $\Psi_b$ by an electromagnetic interaction $H_\gamma$ leading to 3-$\alpha$ states  of energy  
$E$  in the center of mass (c.m.) system. 
As in Ref. \cite{Is13}, a wave function for the reaction is introduced by    
\begin{equation}
\Psi(\boldsymbol{x},\boldsymbol{y})=   \langle \boldsymbol{x},\boldsymbol{y}  \vert \frac{1}{E + \imath \epsilon - H_{3\alpha}}  H_\gamma \vert \Psi_b \rangle,
\label{eq:Psi_def}
\end{equation}
where $H_{3\alpha}$ is the Hamiltonian of the 3-$\alpha$ system, and ($\boldsymbol{x}$, $\boldsymbol{y}$) are Jacobi coordinates,
\begin{equation}
\boldsymbol{x} = \boldsymbol{r}_1 - \boldsymbol{r}_2, 
\qquad
\boldsymbol{y} = \boldsymbol{r}_3 - \frac12 \left( \boldsymbol{r}_1 + \boldsymbol{r}_2 \right), 
\end{equation}
with $\boldsymbol{r}_i$ being the position vector of the $i$-th $\alpha$ particle. 
Eq. (\ref{eq:Psi_def}) is solved by applying the Faddeev three-body formalism  \cite{Fa61} in coordinate space, in which effects of the boson symmetric property of the wave function as well as the long-range Coulomb potential are taken into account properly.
Details of numerical calculations are described in Refs. \cite{Is09,Is13}.

From the solution of Eq. (\ref{eq:Psi_def}), a breakup amplitude 
$F^{(\mathrm{B})}(\hat{\boldsymbol{q}},\hat{\boldsymbol{p}},E_{q})$ is calculated. 
Here, $(\boldsymbol{q}, \boldsymbol{p})$ are Jacobi momenta conjugate to $(\boldsymbol{x}, \boldsymbol{y})$, 
\begin{eqnarray}
\boldsymbol{q} &=& \frac{1}{2} \left( \boldsymbol{k}_1 - \boldsymbol{k}_2 \right),
\cr
\boldsymbol{p}&=& \frac{2}{3}\boldsymbol{k}_3   - \frac{1}{3}\left(\boldsymbol{k}_1 + \boldsymbol{k}_2 \right)
= \boldsymbol{k}_3, 
\end{eqnarray}
where $\boldsymbol{k}_i$ denotes the momentum of the $i$-th $\alpha$ particle in the c.m. system, and $E_q=\frac{\hbar^2}{m_\alpha} \boldsymbol{q}^2$ with $m_\alpha$ being the mass of the $\alpha$-particle.
Note that  $\boldsymbol{q}$ and $\boldsymbol{p}$ satisfy the energy conservation law, 
\begin{equation}
E = \frac{\hbar^2}{m_\alpha} \boldsymbol{q}^2 + \frac{3\hbar^2}{4m_\alpha} \boldsymbol{p}^2, 
\end{equation}
and thus a set of variables ($\hat{\boldsymbol{q}},\hat{\boldsymbol{p}},E_{q}$) is used  to specify kinematical configurations of the three $\alpha$-particles in the c.m. system.

From the amplitude, the number of an event that three $\alpha$-particles take a configuration, 
$\hat{\boldsymbol{q}} \sim \hat{\boldsymbol{q}}+d\hat{\boldsymbol{q}}$, 
$\hat{\boldsymbol{p}} \sim \hat{\boldsymbol{p}}+d\hat{\boldsymbol{p}}$, and 
$E_{q} \sim E_{q}+ dE_{q}$, is calculated by an outgoing flux, 
\begin{equation}
dJ(\hat{\boldsymbol{q}},\hat{\boldsymbol{p}},E_{q}) 
=  
\left\vert F^{(B)}(\hat{\boldsymbol{q}},\hat{\boldsymbol{p}},E_{q}) \right\vert^2
d\hat{\boldsymbol{q}} d\hat{\boldsymbol{p}} dE_{q}.
\label{eq:flux}
\end{equation}
%

{\em Interaction model.}
In this work, the $\alpha$ particle is considered as a boson and every complicacies arising from its nucleon structure are considered to be incorporated in interaction potentials among the $\alpha$-particles, which are usually consisting of two- and three-$\alpha$ potentials. 

I use  the  Ali-Bodmer-D model  \cite{Al66} for the nuclear part of the $\alpha$-$\alpha$ potential along with a point Coulomb potential,
\begin{eqnarray}
V(x) &=& \left( 500 \mathrm{MeV} \hat{P}_{2\alpha,0} + 320 \mathrm{MeV} \hat{P}_{2\alpha,2} \right) e^{-(x/1.40 \mathrm{fm})^2}
\cr
&& - 130\mathrm{MeV} e^{-(x/2.11\mathrm{fm})^2} + \frac{(2e)^2}{x},  
\end{eqnarray} 
where $\hat{P}_{2\alpha,L}$ is a projection operator on the $L$ angular momentum $\alpha$-$\alpha$ state.
(All possible 3-$\alpha$ partial wave states with $L\le4$ are taken into account in the present calculations.)

In addition, a three-body potential (3$\alpha$P) of the following form \cite{Fe96,Is13},
\begin{equation}
W_{3\alpha} = 3 \sum_{J} \hat{P}_{3\alpha,J} W_3^{(J)} 
\exp\left( - {{\sum_{i<j}} \frac{   \left( \boldsymbol{r}_i - \boldsymbol{r}_j\right)^2}{\left(a_3 \right)^2}}\right),
\end{equation}
is introduced, where $\hat{P}_{3\alpha,J}$ is a projection operator on the $J$ angular momentum 3-$\alpha$ state, and the range parameter $a_3$ is chosen to be the same value as in Refs. \cite{Fe96,Is13}. 
The strength parameters are determined to reproduce the energy of the Hoyle state for $J=0$ state and the energy of ${}^{12}$C$(2_1^+)$ bound state for $J=2$ state.
The parameters and calculated energies are summarized in Table \ref{tab:3alphaP-parameters}. 

In view of uncertainties in the interaction of the $\alpha$ particles, I have examined the other $\alpha$-$\alpha$ potential, which is named as AB-A' in Ref. \cite{Is13}, as well as  several other choices for the 3$\alpha$P, whose parameters are determined to reproduce the energies of the Hoyle state and ${}^{12}$C$(2_1^+)$. 
As far as the Hoyle state is concerned, it turns out that calculated  
{distributions of outgoing 3-$\alpha$ particles and density distributions at interior region} 
by these models are essentially the same as those by the present model, which will be shown below.

\begin{table}[tb]
\caption{
Parameters of the 3-$\alpha$ potential and calculated energies and widths of ${}^{12}$C. 
Experimental data are from \cite{Aj90}.
\label{tab:3alphaP-parameters}
}
\begin{ruledtabular}
\begin{tabular}{ccc}
& Model &  Exp.  \\
\hline
$a_3$ (fm) & $\sqrt{3/3.97} \times 3.90$ & \\
$W_3^{(0)}$ (MeV) & -30.95  &  \\
$E[{}^{12}$C$(0_2^+)]$  (MeV) & 0.379177  & 0.3794 \\
$\Gamma_{3\alpha}$ (eV) & 5.8 &  8.3(1.0)\\
 $\Gamma_\gamma$ (meV)  & 2.2 & 3.7(5) \\
$W_3^{(2)}$ (MeV) & -15.3 \\
$E[{}^{12}$C$(2_1^+)]$ (MeV) & -2.83   &-2.8357 \\
\end{tabular}
\end{ruledtabular}
\end{table}

{\em Decay mode of the Hoyle state.}
First, I will investigate the decay of the Hoyle state by calculating the function $\Psi(\boldsymbol{x}, \boldsymbol{y})$  (\ref{eq:Psi_def}) at $E=E[{}^{12}$C$(0_2^+)]$, and thereby the breakup amplitude $F^{(\mathrm{B})}(\hat{\boldsymbol{q}},\hat{\boldsymbol{p}},E_{q})$ and the outgoing flux (\ref{eq:flux}). 
As in the the previous experimental works, the outgoing $\alpha$-particles are ordered  by their energies as    $E_3 \ge E_1 \ge E_2$, where $E_i=\frac1{2m_\alpha}\boldsymbol{k}_i^2$ is the energy of the $i$-th $\alpha$-particle in the c.m. system with $\boldsymbol{k}_i$ being 
\begin{eqnarray}
 \boldsymbol{k}_1  &=& \boldsymbol{q} -  \frac{1}{2}  \boldsymbol{p}, 
\cr
 \boldsymbol{k}_2  &=& -\boldsymbol{q} -  \frac{1}{2}  \boldsymbol{p}, 
\cr
 \boldsymbol{k}_3  &=&    \boldsymbol{p}. 
\label{eq:k_123}
\end{eqnarray}

\begin{figure}[t]
\begin{center}
\includegraphics[width=0.98\columnwidth]{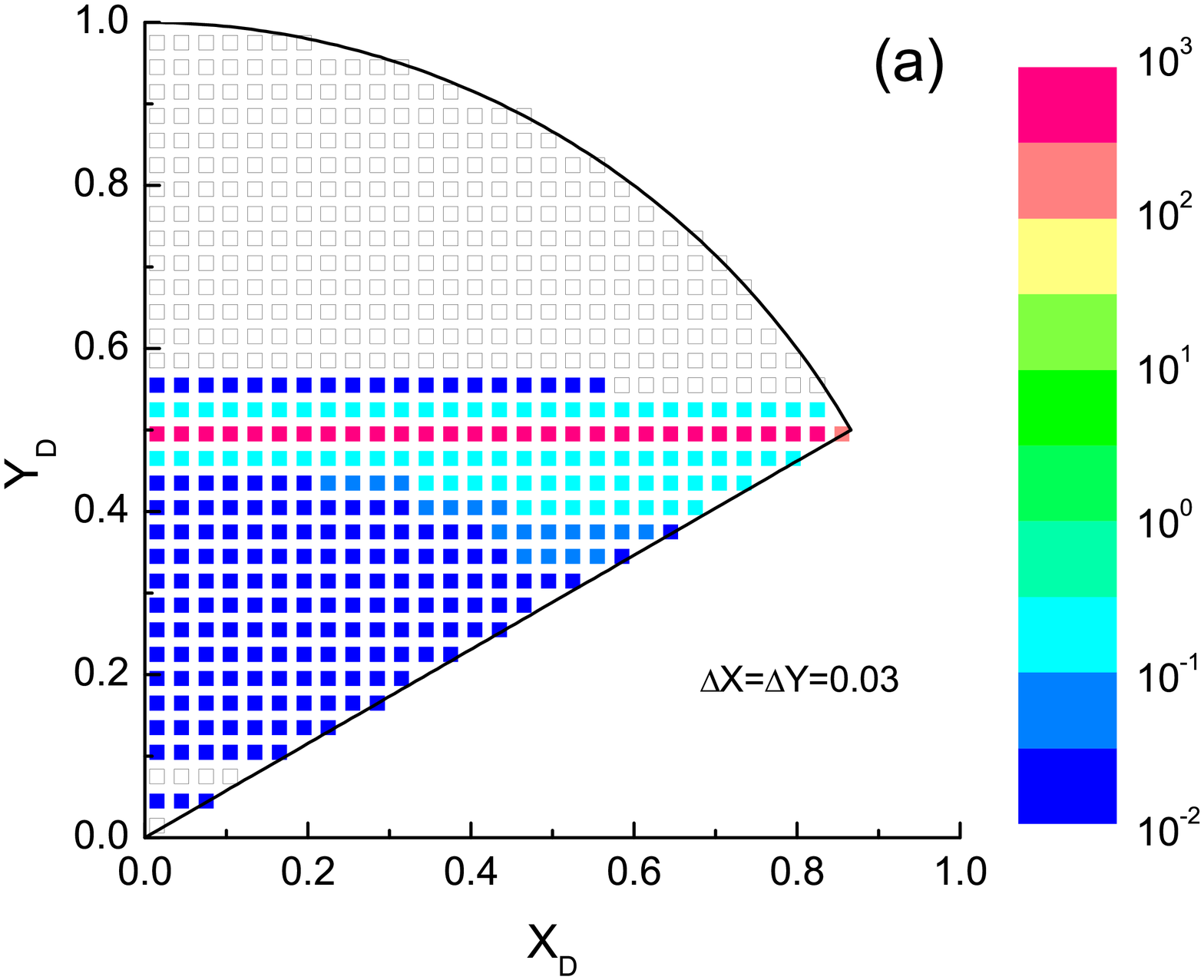}
\includegraphics[width=0.98\columnwidth]{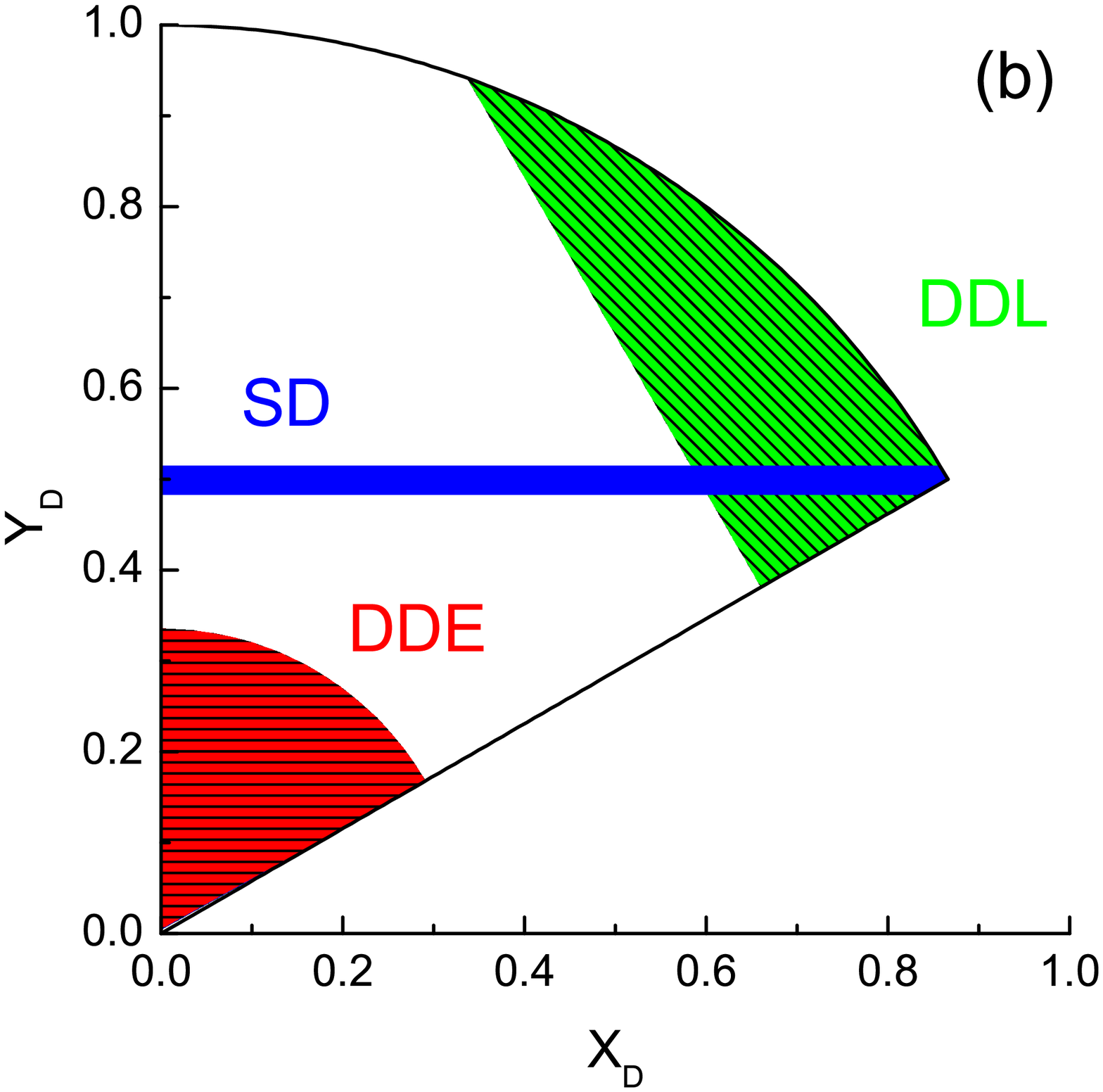} 
\caption{(Color online)
(a) The Dalitz plot for the 3-$\alpha$ decay process of the Hoyle state. 
The number of the events $N(X_D,Y_D)$ is plotted for the variables $X_D$ and $Y_D$ defined in Eq. (\ref{eq:X_D-Y_D}).
(b) The kinematical region for the SD, DDL, and DDE, which are described in the text. 
}
\label{fig:Dalitz-plot}
\end{center}
\end{figure}

In three-body decay reactions, it is convenient to view the distribution of the outgoing particles in the form of Dalitz plot.
Here,  I use  the following two variables,    
\begin{eqnarray}
X_D &=& \sqrt{3} \frac{E_3 + 2 E_1  - E}{E} = -2 \frac{\sqrt{E_{q}(E-E_{q})}}{E} \cos\theta,
\cr
Y_D &=& \frac{3E_3 - E}{E} =  1 - 2 \frac{E_{q} }{E}, 
\label{eq:X_D-Y_D}
\end{eqnarray}
where $\theta$ is the angle between $\hat{\boldsymbol{q}}$ and $\hat{\boldsymbol{p}}$.
In the $X_D-Y_D$ plane, every events with  $E_3 \ge E_1 \ge E_2$ are located in the area that $0 \le \sqrt{X_D^2 +Y_D^2} \le 1$ and $ \pi/6 \le \arctan(Y_D/X_D) \le \pi/2$. 
The area is divided to cells of the size $\Delta X_D \times \Delta Y_D$ and the number of events $N(X_D,Y_D)$ is calculated by integrating the flux $dJ(\hat{\boldsymbol{q}},\hat{\boldsymbol{p}},E_q)$ in the cell, where $(X_D, Y_D)$ is the position of the center of the cell.  
With setting the total number of the events to be $2\times10^4$ as  ones in recent experiments  \cite{Ra13,It14},  the $N(X_D,Y_D)$ for $\Delta X_D=\Delta Y_D=0.03$ is displayed in Fig. \ref{fig:Dalitz-plot} (a).   
In this plot, there is a sharp ridge at $Y_D \sim 1/2$ corresponding to two $\alpha$ particles (1 and 2) being  the $^8$Be$(0_1^+)$ state.
(Note that $E[{}^8$Be$(0_1^+)] \sim E/4$.)
The number of the events for $0.48 \le Y_D \le 0.51$ ($\Delta E_{q} \sim  $ 6 keV), which should be assigned as the SD mode, is about 99.9 \% of the total.

The above calculation demonstrates that the SD contribution exceeds 99 \% of the total events.
Of the rest events,  which are assigned as a direct decay,  two different decay modes have induced interests: a decay with a linear chain like configuration (DDL) and one with three $\alpha$-particles with equal energy in the c.m. (DDE).
Both modes are kinematically defined as follows.
In the DDL mode, one of the three $\alpha$-particles, the particle 2 in this case, stays at the c.m. of the system,  i.e., $E_2=0$.
The DDE mode is  defined as $E_{\mathrm{rms}} = 0$, where $E_{\mathrm{rms}} = \sqrt{\langle E_{\alpha}^2 \rangle - \langle E_{\alpha} \rangle^2 }$, $\langle E_{\alpha}^2 \rangle = \frac{1}{3} \sum_{i=1,3} E_i^2$, and $\langle E_{\alpha} \rangle = \frac{1}{3} \sum_{i=1,3} E_i = \frac{1}{3}E$.
(Note that $E_{\mathrm{rms}} = \frac{E}{3\sqrt{2}} \sqrt{X_D^2 +Y_D^2}$.)

In actual calculations,  I evaluate the contributions of the DDL and DDE modes by setting $\delta E_{\mathrm{DDL}}$ and $\delta E_{\mathrm{DDE}}$ as decent values and then integrating the flux $dJ(\hat{\boldsymbol{q}},\hat{\boldsymbol{p}},E_q)$ with conditions that $E_{2} \le \delta E_{\mathrm{DDL}}$ and  $E_{\mathrm{rms}} \le \delta E_{\mathrm{DDE}}$, respectively. 
Regions for the DDL and DDE  with $\delta E_{\mathrm{DDL}} = \delta E_{\mathrm{DDE}} = 30$ keV in the $X_D-Y_D$ plane are displayed in Fig. \ref{fig:Dalitz-plot} (b) together with the region for the SD mode. 
Note that there is an overlapped region between SD and DDL, and then the SD contribution is excluded in evaluating the DDL contribution. 
These procedures give  0.03 \% for the DDL contribution,  and 0.005 \% for the DDE. 
It is noted that  contributions of DDL and DDE modes stay unchanged even when calculated at energies shifted from $E[{}^{12}$C$(0_2^+)]$ by a few times of $\Gamma_{3\alpha}$. 
Thus the direct decay modes at energies around the Hoyle state are the same  as those at the resonance energy.

\begin{figure}[t]
\begin{center}
\includegraphics[width=0.96\columnwidth]{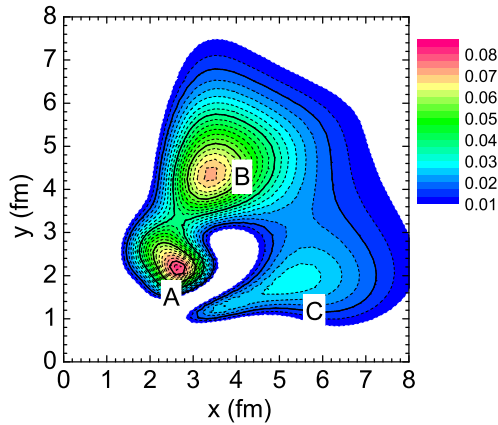} 
\caption{(Color online)
Contour plot of the density distribution  $\rho(x,y)$ of the Hoyle state.
}
\label{fig:density-ab-d-w}
\end{center}
\end{figure}

{\em Structure of the Hoyle state.}
The function $\Psi (\boldsymbol{x},\boldsymbol{y})$ at the resonance energy has a concentration of the amplitude at interior region. 
In Fig. \ref{fig:density-ab-d-w}, the density distribution  
\begin{equation}
\rho(x,y) = x^2 y^2 \int d\hat{\boldsymbol{x}} d\hat{\boldsymbol{y}}   
\vert \Psi(\boldsymbol{x},\boldsymbol{y})\vert^2, 
\end{equation}
calculated at the Hoyle state energy is plotted.
Note that $\Psi (\boldsymbol{x},\boldsymbol{y})$ is not square normalizable, and thus it is artificially normalized within the region, $0 \le x \le x_\mathrm{max}$, and $0 \le y \le y_\mathrm{max}$  with $x_\mathrm{max} = y_\mathrm{max} = 12$ fm.
The density has three distinct local peaks denoted by A, B, and C in the figure, which are located at $(x, y) \sim (2.5~ \mathrm{fm}, 2.2~ \mathrm{fm})$, $(3.3~ \mathrm{fm}, 4.2~ \mathrm{fm})$, and $(5.3~ \mathrm{fm}, 1.9~ \mathrm{fm})$, respectively.
Similar peak structure is observed in calculations of Refs. \cite{Ng13,Va12}.

To reveal the 3-$\alpha$ structure more precisely,  I calculate an intrinsic density distribution in a body-fixed frame $\rho_{\mathrm{bf}}(x,y,\Theta)$, where $\Theta$ is the angle between $\hat{\boldsymbol{x}}$ and $\hat{\boldsymbol{y}}$. 
By defining Euler angles $\boldsymbol{\Omega}$ associated with a rotation to a body-fixed frame ($XYZ$), in which the $Z$-axis is chosen along the vector $\boldsymbol{x}$ and the $XZ$-plane on the plane of 3-$\alpha$,
the intrinsic density is calculated  as 
\begin{equation}
\rho_{\mathrm{bf}}(x,y,\Theta) = x^2 y^2 \int d\boldsymbol{\Omega}   
\vert \Psi(\boldsymbol{x},\boldsymbol{y})\vert^2.
\end{equation}

\begin{figure}[t]
\begin{center}
\includegraphics[width=0.8\columnwidth]{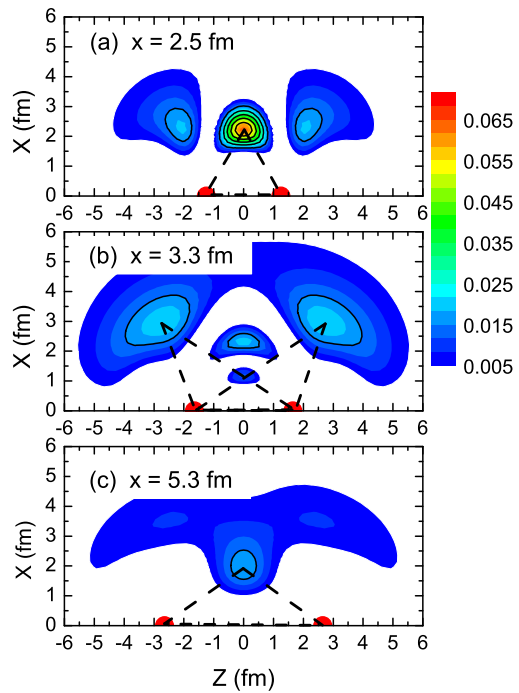} 
\caption{(Color online)
Contour plots of the intrinsic density distribution $\rho_{\mathrm{bf}}$ of the Hoyle state for fixed position of two $\alpha$-particles.
In figure (a), (b), and (c), two $\alpha$-particles are positioned separating by 2.5 fm, 3.3 fm, and 5.3 fm, respectively.
The positions of the two $\alpha$-particles are denoted by the red points in each figure.  
}
\label{fig:yz-yx}
\end{center}
\end{figure}

When two $\alpha$-particles are fixed in a distance $x$, the position of the third $\alpha$-particle on $XZ$-plane is given by $(X, Z) = (y\sin\Theta, y\cos\Theta)$. 
In Fig. \ref{fig:yz-yx},  the $\rho_{\mathrm{bf}}(x,y,\Theta)$ densities with fixing $x$ to be the peak positions of $\rho(x,y)$, namely (a) $x=2.5$ fm, (b) 3.3 fm, and (c) 5.5 fm, are plotted. 
As a reference, an equilateral triangle of side length 2.5 fm is drawn by dashed-line in Fig. \ref{fig:yz-yx} (a).
Also, isosceles triangles with two equal sides of length 3.3 fm and the third side length being 5.3 fm  are drawn in  Figs. \ref{fig:yz-yx} (b) and (c).
The figures show that the peak A corresponds to the configuration of the equilateral triangle, and  that the peaks B and C correspond to a bent-arm configuration, in which three $\alpha$-particles compose the isosceles triangle.
Because of the symmetric property of the wave function, a bent-arm configuration appears at three points in the density distribution Figs. \ref{fig:yz-yx} (b) and (c).

Since each peak is associated with wide slopes, three $\alpha$-particles may take each triangle configuration rather loosely. 
The probability to find $\alpha$-particles taking the equilateral triangle configuration is estimated by integrating the density $\rho(x,y)$  over a domain of a square, 1.5 fm on a side, around the peak A.
This gives about 10 \%. 
Similar procedure for the peak B (C) gives about 20 \% (10\%).  
Therefore, the Hoyle state has a mixed configuration of the equilateral triangle with probability 10 \% 
and the bent-arm with 30 \%.

Since the Hoyle state is a resonant state, whose wave function does not decay exponentially, observables such as a radius are not well defined. 
However, because of a resonant character it may simulate a bound state if  one restrict the wave function within the interior region where the wave function is normalized.
Here,  the root mean square radius  is calculated by using the following formula,
\begin{equation}
R_\textrm{rms} 
= \sqrt{R_\alpha^2 + \frac{1}{6} \langle x^2 \rangle  + \frac{2}{9} \langle y^2\rangle},
\end{equation}
where $R_\alpha=1.47$ fm, and $\langle x^2 \rangle$ and $\langle y^2\rangle$ are expectation values of $x^2$ and $y^2$ for the  wave function normalized and integrated in the region $0 \le x \le x_\mathrm{max}$ and $0 \le y \le y_\mathrm{max}$. 
Calculated values of  $R_\textrm{rms}$ depend on the choice of $x_\mathrm{max}$ and $y_\mathrm{max}$.
It turns out that calculated values of $R_\textrm{rms}$ with $x_\mathrm{max}=y_\mathrm{max}$= 8.0 to 15.0 fm are well fitted by  
\begin{equation}
R_\textrm{rms} (x_\mathrm{max}) = 3.43 - 4.87\times 0.70^{x_\mathrm{max}}, 
\end{equation}
which gives asymptotically $R_\textrm{rms} =$ 3.43 fm.
This rather large radius is consistent with calculations given in Refs. \cite{Ch07,Ka07}.

{\em Summary.}
%
Decay modes and the structure of the Hoyle state are studied in the 3-$\alpha$ model by calculating  3-$\alpha$  breakup reactions of the ${}^{12}$C$(2_1^+)$  state by the E2 photon. 
Since little dependence of the results on the choice of $\alpha$-particle interaction models was found, calculations with the Ali-Bodmer-D $\alpha$-$\alpha$ potential together with a 3-$\alpha$ potential  are presented. 
The density distribution at interior region of the Hoyle state has peaks corresponding to the configuration of  the equilateral triangle of side length 2.5 fm  and that of the isosceles triangle with two equal sides of length 3.3 fm and the third side length being 5.3 fm. 
The latter corresponds to the bent-arm configuration.
Both configurations have wide slopes, which means the Hoyle state is a weak mixture of these configurations. 
On the other hand, such a structure does not influence the configuration of the outgoing three $\alpha$-particles, which is dominated by the sequential decay process through the ${}^8$Be$(0_1^+)$.  
Two-body interaction of two $\alpha$-particles plays an important role when $\alpha$-particles spread.


\end{document}